\documentstyle[12pt,epsbox]{article}
\newcommand\beq{\begin{eqnarray}}
\newcommand\eeq{\end{eqnarray}}

\newcommand\la{\langle}
\newcommand\ra{\rangle}

\begin{document}

\Large

\centerline{\bf $A_{LT}$ in the polarized Drell-Yan process} 
\centerline{\bf at RHIC and HERA 
energies}

\normalsize

\vspace{1cm}

\centerline{Y. Kanazawa, Yuji Koike and N. Nishiyama}
\centerline{\it  Graduate School of Science and Technology,
Niigata University}
\centerline{\it Ikarashi, Niigata 950-21, Japan}

\vspace{1cm}

\centerline{\bf Abstract}

We present a leading order (LO) 
estimate for the longitidinal-transverse 
spin asymmetry ($A_{LT}$) in the nucleon-nucleon polarized
Drell-Yan process at RHIC and HERA-$\vec{N}$ energies
in comparison with $A_{LL}$ and $A_{TT}$.
$A_{LT}$ receives contribution from $g_1$, the transversity
distribution $h_1$, and the twist-3 distributions $g_T$
and $h_L$.  For the twist-3 contribution we use
the bag model prediction evolved to a high energy scale by
the large-$N_c$ evolution equation.  We found that
$A_{LT}$ (normalized by the asymmetry in the parton level)
is much smaller than the corresponding $A_{TT}$.
Twist-3 contribution given by the bag model 
also turned out to be negligible.

\vspace{1cm}

\noindent
[keywords:  Drell-Yan, longitudinal-transverse spin asymmetry,
chiral-odd distribution, twist-3 effect]

\newpage

\setcounter{equation}{0}
\renewcommand{\theequation}{\arabic{equation}}

\vskip 1cm

The nucleon-nucleon scattering provides us with a new opportunity
to probe nucleon's internal structure.
In particular, polarized Drell-Yan
lepton pair production opens a window toward
new types of spin dependent parton distibutions -- chiral-odd
distributions $h_1(x,\mu^2)$
and $h_L(x,\mu^2)$ which can not be measured by the deep
inelestic lepton-nucleon scatterings\,\cite{RS,AM,JJ,CPR}.
There are 
three kinds of double
spin asymmetries in the nucleon-nucleon polarized Drell-Yan process:
They are $A_{LL}$ (collision between the  
longitudinally polarized nucleons), $A_{TT}$ (
collision between the 
transversely polarized nucleons), and $A_{LT}$ (longitudinal
versus transverse).
The experimental data on these asymmetries will presumably
be reported by RHIC at BNL and HERA-$\vec{N}$ at DESY.
By now
several reports are already available for the
estimate of $A_{LL}$ and
$A_{TT}$
in the leading order (LO) 
and the next-to-leading order (NLO)
level\,\cite{VW,Kama96,BCD,Geh,Kama97,MSSV}. 
The purpose of this short note is to present a first
estimate on $A_{LT}$ in comparison with $A_{LL}$ and $A_{TT}$
at RHIC and HERA energies in the LO QCD. 
Our interest in $A_{LT}$ is amplified by the fact that
it receives the twist-3 contribution as a leading contribution
(although it is proportional to $1/Q$
with a hard momentum $Q$), giving a possibility
of seeing quark-gluon correlation in hard processes.\footnote{
Other recommended twist-3 observables are, for example,
the spin structure function $g_2(x,Q^2)$ in the
transversely polarized deep ineleastic scattering\,\cite{Jg2} and
the single transverse spin asymmetry $A_N$ 
in the nucleon-nucleon direct photon production\,\cite{QS}.}

We first recall the parton distributions relevant to
these asymmetries.
For the nucleon moving in the positive $\hat{\bf e}_3$ direction,
the parton distibutions of the nucleon defined at a
factorization scale $\mu^2$
are given 
by the following lightcone correlation functions in the nucleon
($z^2=0,\ z^+=0,\ \vec{\bf z}_\perp =\vec{\bf 0}$):
\beq
& &P^+\int{dz^- \over 2\pi} e^{i xP\cdot z}\la PS|\bar{\psi}
(0)\gamma_\mu \psi(z)|_\mu |PS \ra  = 2 f_1(x,\mu^2)P_\mu,
\label{eq1}\\
& &P^+\int{dz^- \over 2\pi} e^{ixP\cdot z}\la PS|\bar{\psi}
(0)\gamma_\mu \gamma_5 \psi(z)|_\mu |PS \ra  
=2\left[ g_1(x,\mu^2)P_\mu (S\cdot n) + g_T(x,\mu^2) 
S_{\perp\mu} \right],
\label{eq2}\\
& &P^+\int{dz^- \over 2\pi} e^{ixP\cdot z}\la PS|\bar{\psi}
(0)\sigma_{\mu\nu} i\gamma_5 \psi(z)|_\mu |PS \ra 
\nonumber\\
& & \qquad=
2\left[ h_1(x,\mu^2)\left( S_{\perp\mu}P_\nu -  S_{\perp\nu}P_\mu \right)/M
+h_L(x,\mu^2)M \left( P_\mu n_\nu - P_\nu n_\mu \right) (S\cdot n)
\right],
\label{eq3}
\eeq
where $|PS\ra$ denotes the nucleon (mass $M$) state with the four momentum
$P$ and the spin $S$ ($P^2 =M^2, S^2=-M^2, P\cdot S =0$), 
and a light-like vector $n$ with its only nonzero component $n^-$
is introduced by the relation
$P\cdot n=1$.
$S^\mu$ is decomposed as $S^\mu=(S\cdot n) P^\mu -M^2 (S\cdot n) n^\mu + 
S_\perp^\mu$ with $P\cdot S_\perp= n\cdot S_\perp=0$.  
In (\ref{eq1})-(\ref{eq3}), 
the gauge link operators which ensure gauge invariance
are suppressed for simplicity.
We remind that in the infinite momentum frame ($P^+\to \infty$) the
coefficients of $f_1$, $g_1$ and $h_1$ are of $O(P^+)$ (twist-2),
those of $g_T$ and $h_L$ are of $O(1)$ (twist-3), and 
the $O(1/P^+)$ contributions (twist-4) are ignored
in the right hand side of (\ref{eq1})-(\ref{eq3}).
Note also that $g_1$ and $h_L$ are associated with
the longitudinal polarization, and $h_1$ and $g_T$
are associated with the transverse polarization of the nucleon.
The above distribution functions $g_{1,T}$ 
and $h_{1,L}$ {\it etc} 
are defined for each quark and anti-quark flavor
$\psi = \psi^a$ ($a=u, d, s, \bar{u}, \bar{d}, \bar{s},..$) 
and have support $-1< x <1$.  
They represent distributions of a quark (or anti-quark) 
carrying the momentum component $k^+=xP^+$
in the nucleon. 
A quark and an anti-quark distributions are
related as 
$f_1^a(-x)=-f_1^{\bar{a}}(x)$,
$g_{1,T}^a(-x)=g_{1,T}^{\bar{a}}(x)$,
$h_{1,L}^a(-x)=-h_{1,L}^{\bar{a}}(x)$.
The 
three twist-2 distributions 
have a simple parton model interpretation, and they can be written as
$f_1(x)=q_+(x)+q_-(x) =q_\uparrow (x) + q_\downarrow (x)$, 
$g_1(x)=q_+(x)-q_-(x)$, and
$h_1(x)=q_\uparrow (x) - q_\downarrow (x)$. 
Here $q_+(x)$ ($q_-(x)$) represents a density of a quark
with 
its helicity parallel (anti-parallel) to the nucleon spin
in the longitudinally polarized nucleon.
Likewise, $q_\uparrow(x)$ ($q_\downarrow(x)$) 
represents a density of a quark with 
its polarization parallel (anti-parallel) to the nucleon spin
in the transversely polarized nucleon.
Therefore $h_1$ is called transversity distribution\,\cite{JJ}. 
For nonrelativistic quarks, $h_1(x)=g_1(x)$.  Nucleon models
suggest $h_1(x)$ is not so different from $g_1(x)$
at a low energy scale\,\cite{JJ,PP}.

The twist-3 distributions $g_T$ and $h_L$ can be decomposed into
the twist-2 contribution  
and the ``purely twist-3'' contribution:
\beq
g_{T}(x,\mu^2) &=& \int_{x}^{1} dy\frac{g_{1}(y,\mu^2)}{y}  + 
\widetilde{g}_{T}(x,\mu^2),
\label{gTWW}\\
h_{L}(x,\mu^2) &=& 2x\int_{x}^{1}dy \frac{h_{1}(y,\mu^2)}{y^2}  
+ \widetilde{h}_{L}(x,\mu^2).\qquad
\label{hLWW}
\eeq
The purely twist-3 pieces
$\widetilde{g}_T$ and
$\widetilde{h}_L$ can be written as 
quark-gluon-quark correlator on the lightcone
using QCD equation of motion
\cite{g2,JJ,BBKT}.  
In the following we call the 
first terms in (\ref{gTWW}) and (\ref{hLWW})
$g_T^{WW}(x,\mu^2)$ and $h_L^{WW}(x,\mu^2)$ (Wandzura-Wilczek parts)
respectively.

With these definition for the parton distributions, 
we can write down the expression for the double spin asymmetries, 
$A_{LL}$, $A_{TT}$\,\cite{RS}, $A_{LT}$\,\cite{JJ}, 
in the polarized Drell-Yan process.  In LO QCD, 
there is some arbitrariness in choosing the factorization
scale $\mu^2$ in each distribution.  In this paper, we set 
$\mu^2=Q^2$, the squared invariant mass of the lepton pairs, 
in calculating the asymmetries.  By this
choice the LO double asymmetries are given by
\beq
A_{LL} &=& \frac{\sigma (+,+)-\sigma (+,-) }
                {\sigma (+,+)+\sigma (+,-) }
       =  \frac{\Sigma_{a} e_{a}^2
                   g_{1}^a(x_1,Q^2)g_{1}^{\bar{a}}(x_2,Q^2)}
               {\Sigma_{a} e_{a}^2 f_{1}^a(x_1,Q^2)f_{1}^{\bar{a}}(x_2,Q^2)},
\label{ALL}\\[5pt]
A_{TT} &=& \frac{\sigma (\uparrow,\uparrow)-\sigma (\uparrow,\downarrow) }
                {\sigma (\uparrow,\uparrow)+\sigma (\uparrow,\downarrow) }
       = a_{TT} \frac{\Sigma_{a} e_{a}^2
                   h_{1}^a(x_1,Q^2)h_{1}^{\bar{a}}(x_2,Q^2)}
               {\Sigma_{a} e_{a}^2 f_{1}^a(x_1,Q^2)f_{1}^{\bar{a}}(x_2,Q^2)},
\label{ATT}\\[5pt]
A_{LT} &=& \frac{\sigma (+,\uparrow)-\sigma (+,\downarrow) }
                {\sigma (+,\uparrow)+\sigma (+,\downarrow) }
       = a_{LT} \frac{\Sigma_{a} e_{a}^2
                   \left [g_{1}^a(x_1,Q^2)x_2g_{T}^{\bar{a}}(x_2,Q^2)
                        + x_1h_{L}^a(x_1,Q^2)h_{1}^{\bar{a}}(x_2,Q^2) \right]}
               {\Sigma_{a} e_{a}^2 f_{1}^a(x_1,Q^2)f_{1}^{\bar{a}}(x_2,Q^2)},
\nonumber\\
\label{ALT}
\end{eqnarray}
where $\sigma(S_1,S_2)$ represents the Drell-Yan cross section
with the two nucleon's spin $S_1$ and $S_2$, 
$e_a$ represent the electric charge of the quark-flavor
$a$ and the summation is over all quark and anti-quark flavors: 
$a=u,d,s,\bar{u},\bar{d},\bar{s}$, ignoring heavy quark 
contents ($c,b,\cdots$) 
in the nucleon.
The variables $x_1$ and $x_2$ refer to
the momentum fractions of the partons coming from
the two nucleons ``1'' and ``2'', respectively.
In $A_{LT}$, the nucleon ``1'' is longitudinally
polarized and the nucleon ``2'' is transversely polarized.
In (\ref{ATT}) and (\ref{ALT}), $a_{TT}$ and $a_{LT}$ represent
the asymmetries in the parton level defined as
\beq
a_{TT}  &=&  \frac{{\rm sin}^2\theta\, 
{\rm cos}2\phi}{1+{\rm cos}^2\theta},
\label{aTT}\\
a_{LT}  &=&  \frac{M}{Q} \frac{2\,{\rm sin}2\theta\, 
{\rm cos}\phi}{1+{\rm cos}^2\theta},
\label{aLT}
\eeq
where $\theta$ is the
polar angle of the virtual photon in the center of mass system 
with respect to the beam direction and $\phi$
represents its azimuthal angle with resepect to the transverse spin.
We note that $A_{LL}$ and $A_{TT}$ receive contribution only from
the twist-2 distributions, while $A_{LT}$ is proportional to the twist-3 
distributions
and hence $a_{LT}$ is suppressed by a factor $1/Q$.

So far there has been much accumulation of experimental data 
on $f_1$ and $g_1$, and they have been parametrized in the 
NLO level in the literature. (\cite{GRV,MRS,CTEQ} 
for $f_1$ and \cite{GS,GRSV,ABFR} for $g_1$)
Although  
these distributions can explain available experimental data,
there still remains 
some uncertainties in the parametrizations
especially for $g_1$.
For future references, we use for $g_1$ the
LO parametrization (standard scenario)
by Glu\"ck-Reya-Stratmann-Vogelsang (GRSV)\,\cite{GRSV}
and the LO model-A of Gehrmann and Stirling (GS)\,\cite{GS} for our estimate.
In both cases, 
we consistently use the LO parametrization for $f_1$ by 
Glu\"ck-Reya-Vogt\,\cite{GRV}.  
For $h_1$, $g_T$ and $h_L$ no experimental
data is available up to now and we have to rely on some theoretical
postulates. 
Here we assume $h_1(x,\mu^2)=g_1(x,\mu^2)$ at a low energy scale
($\mu^2=0.23$ GeV$^2$ for GRSV
and $\mu^2=1$ GeV$^2$ for GS)
as has been suggested by some low energy nucleon models\,\cite{JJ,PP}
and has been used for the estimate of $A_{TT}$\,\cite{BCD}.
This assumption also fixes $g_T^{WW}$ and $h_L^{WW}$.
For the purely twist-3 parts $\widetilde{g}_T$ and
$\widetilde{h}_L$ we employ the bag model results at a low energy 
scale.  In particular, we set the strangeness contributions
to the purely twist-3 contributions equal to zero.
By these boundary conditions for $h_1$, $g_T$ and $h_L$
at a low energy side
and applying the relevant $\mu^2$ evolution to them,
we can estimate $A_{LT}$.

For the LO prediction of the asymmetries, 
we need LO 
$\mu^2$ evolution for each distribution.  
The twist-2 distributions obey simple DGLAP equation.
The complete LO $\mu^2$ evolution of the twist-3 distributions
has been derived
by several different approaches for
$\widetilde{g}_T$\,\cite{g2} and for $\widetilde{h}_L$\,\cite{KT}.
It has been also proved that at large $N_c$ their $\mu^2$-dependence 
can be described by a simple DGLAP evolution equation
similarly for the twist-2 distributions 
and the correction due to the finite value of $N_c$ is 
of $O(1/N_c^2)\sim 10$ \% level\,\cite{ABH,BBKT}.
Since the complete evolution equations of $\widetilde{g}_T$ 
and $\widetilde{h}_L$
are quite complicated and 
not practically usefull, we apply the large-$N_c$
evolution to the bag model results\,\cite{KK,St}.

The double spin asymmetries are the functions of
the square of the center-of-mass energy
$s=(P_1 + P_2)^2$ ($P_1$ and $P_2$ are the four momenta of the
two nucleons), 
the squared invariant mass of
the lepton pair 
$Q^2=(x_1 P_1 + x_2 P_2)^2 = x_1 x_2 s$ ($M^2 << Q^2$) and the Feynman's 
$x_F={2q_3\over \sqrt{s}}=x_1-x_2$.  
Using these variables, momentum fractions of
each quark and anti-quark in (\ref{ALL})-(\ref{ALT}) can be written
as
\beq
x_1 = {1\over 2}\left( x_F +\sqrt{x_F^2 + {4Q^2\over s}}\right),\qquad
x_2 = {1\over 2}\left( -x_F +\sqrt{x_F^2 + {4Q^2\over s}}\right).
\label{xf}
\eeq

Since the twist-3 effect is one of our main interest,
we showed in Fig. 1 $g_T(x,\mu^2)$ and $h_L(x,\mu^2)$ 
for the $u$-quark at $\mu^2=1$ GeV$^2$.
Figure 1(a) shows $g_T^{WW}(x,\mu^2)$ with two parametrizations 
for $g_1$
(GRSV and GS) and the bag model prediction for $\widetilde{g}_T(x,\mu^2)$
obtained by assuming the bag scale is $\mu_{bag}^2 = 0.081$ and $0.25$ 
GeV$^2$\,\cite{KK,St}.
Although the $u$-quark distribution for $\widetilde{g}_T$
contains flavor-singlet contribution which mixes with the gluon distribution,
we ignored the mixing and 
applied the large-$N_c$ $\mu^2$ evolution, since
the $\mu^2$ evolution for the singlet part can not be described by
a simple evolution equation. 
The correction due to the mixing is expected to be 
at most of order of 10 \%,
which is irrelevant in the present rough estimate of $A_{LT}$
as we will see later.
Likewise, Fig. 1(b) shows
$h_L^{WW}(x,\mu^2)$ and $\widetilde{h}_L(x,\mu^2)$
at $\mu^2=1$ GeV$^2$.
Because of chiral-odd nature of $h_L$, $\widetilde{h}_L$
does not mix with the gluon distribution and the large-$N_c$
evolution is more reliable than for $\widetilde{g}_T$.
One sees from Fig. 1, the GRSV and GS distributions
give rise to different twist-2 contributions $g_T^{WW}$ and 
$h_L^{WW}$ at $x<0.2$.   
It has been shown in \cite{KK,St} that 
at high $\mu^2$ the bag model gives
small $\widetilde{g}_T$ and $\widetilde{h}_L$,  
and $g_T$ and $h_L$ are dominated by  $g_T^{WW}$ and 
$h_L^{WW}$.  Figure 1 reflects this tendencey already at $\mu^2=1$ GeV$^2$.

Figure 2 shows the three asymmetries normalized by
the asymmetries in the parton level, $\widetilde{A}_{LL}=-A_{LL}$,
$\widetilde{A}_{TT}=-A_{TT}/a_{TT}$, $\widetilde{A}_{LT}=-A_{LT}/a_{LT}$
using the GRSV distribution for the
twist-2 distributions.  They are plotted as a function of $x_F$
for fixed values of 
$Q=\sqrt{Q^2}$ ($=8,\ 10$ GeV) and $\sqrt{s}$ ($=50,\ 200$ GeV),
which are within or 
close to the planned RHIC and HERA-$\vec{N}$
kinematics. ($50\ {\rm GeV}< \sqrt{s}< 500\ {\rm GeV}$ for RHIC,
and $\sqrt{s}=39.2$ GeV for HERA-$\vec{N}$.)  
Figure 3 shows the same quantities but with the GS
distributions for the twist-2 distributions.  In Figs. 2 and 3, 
$\widetilde{A}_{LL}$ and 
$\widetilde{A}_{TT}$ are symmetric with respect to $x_F=0$, while
$\widetilde{A}_{LT}$ is not symmetric 
as is obvious from the kinematics.  In general all these asymmetries are
larger for larger $Q^2/s$.
One sees from these figures that even for 
$\widetilde{A}_{LL}$ and $\widetilde{A}_{TT}$
the GRSV and GS parton distributions give completely different
results.  (See also \cite{Geh}.) :  Their $x_F$-dependence
is mostly opposite.
Comparing the curves with $\sqrt{s}= 50$ GeV and 200 GeV in Fig. 2,
the relative magnitude of $\widetilde{A}_{LL}$ and 
$\widetilde{A}_{TT}$ is reversed.
The GS distribution for $g_1$ gives negative $\widetilde{A}_{LL}$ 
in some range of $x_F$ at $\sqrt{s}=50$ GeV.
$\widetilde{A}_{LT}$ with only the twist-2 contributions in $g_T$ and $h_L$
are shown by solid lines in Figs. 2 and 3.
They are
typically 5 to 10 times smaller than $\widetilde{A}_{LL}$ and
$\widetilde{A}_{TT}$ except when $\widetilde{A}_{LL}$ 
changes sign in Fig. 3.  
$\widetilde{A}_{LT}$ with complete $g_T$ and $h_L$ is shown by
the short dash-dot ($\mu_{bag}^2=0.25$ GeV$^2$) and the dotted
($\mu_{bag}^2=0.081$ GeV$^2$) lines in these figures.
Since large $|x_F|$ corresponds to small $x_1$ or $x_2$ (see (\ref{xf})),
and the bag model prediction for the distribution function
becomes unreliable in the small-$x$ region, we only plotted these lines
for the region $x_1, x_2 > 0.07$\,\cite{Jaffe75}.   As can be seen from
Figs. 2 and 3, the purely twist-3 contribution brings only tiny correction
to $\widetilde{A}_{LT}$.   Larger value of the bag scale
$\mu^2_{bag}$ would not make it appreciably larger.

Figures 4 and 5 show the $Q^2$ dependence of the asymmetries
at $x_1=x_2=Q/\sqrt{s}$ ($x_F=0$) with $\sqrt{s}=50,\ 200$ GeV,
using two distributions.  These figures also show the features noted above.

The reason for the smallness of $\widetilde{A}_{LT}$
is the presence of the factors $x_1$ or $x_2$ in (\ref{ALT}).
In the kinematic range considered either $x_1$ or $x_2$ (or both)
take very small values.  
If it were not for those factors, $\widetilde{A}_{LT}$
would be comparable to $\widetilde{A}_{LL}$ and $\widetilde{A}_{TT}$.  
We remind in passing 
that what is measured experimentaly is $A_{LT}$ itself
which receives the suppression factor $M/Q$ from $a_{LT}$. 

Several comments are in order here.  Our estimate of $A_{LT}$
is based on the assumption $h_1^a(x,\mu^2)=g_1^a(x,\mu^2)$ 
at low $\mu^2$.  There are several model independent 
constraints among LO twist-2 parton distributions:
$f_1^a(x,\mu^2) \geq |g_1^a(x,\mu^2)|$, $f_1^a(x,\mu^2)\geq |h_1^a(x,\mu^2)|$
and the Soffer's 
inequality $f_1^a(x,\mu^2)+g_1^a(x,\mu^2) 
\geq 2 |h_1^a(x,\mu^2)|$\,\cite{Soffer}. 
Since the first inequality is satisfied by the GRSV and GS distributions
at $\mu^2 > 1$ GeV$^2$, the second one is also satisfied
by our input for $h_1^a$ and its $\mu^2$ evolution.  
On the other hand, our input assumption
$h_1^a(x,\mu^2)=g_1^a(x,\mu^2)$ at low $\mu^2$ may violate
the Soffer's inequality even at relatively high $\mu^2$
for some quark or anti-quark flavors 
for which $g_1^a(x,\mu^2) \approx -f_1^a(x,\mu^2)$ at low $\mu^2$. 
For the case of the GRSV distribution, our assumption violates
the inequality for the $\bar{u}$, $d$, $s$ and $\bar{s}$ distributions 
even at $\mu^2 > 1$ GeV$^2$ in the large $x$ region.  
For the case of GS distribution,
we found that the $d$ quark distribution violates the inequality
at $x>0.5$ around $\mu^2=4$ GeV$^2$.
This may be ascribed not only to our exact setting of
$h_1^a(x,\mu^2)=g_1^a(x,\mu^2)$ at low $\mu^2$ but also
to the uncertainty in $g_1^a(x,\mu^2)$, in particular, the poor
knowlegde on sea distributions at $x>0.1$.
However, the violation occurs only in the region where
the absolute magnitude of the distribution functions is extremely
small.  We thus expect that 
the effect of the violation to the asymmetries is not so serious numerically.
Especially, we believe that the relative magnitude between $A_{LT}$
and $A_{TT}$ is relatively immune to this constraint.

The authors of \cite{MSSV} calculated $A_{TT}$ from a different point of
view.  They 
determined the input $h_1$ so that it saturates
the Soffer's inequality
at a low energy scale, taking advantage 
of the fact that the inequality is maintained at higher $\mu^2$ by the QCD
evolution, and claimed that they estimated the upper bound of $A_{TT}$.  
However, they assumed the inequality for each {\it valence}
and {\it sea} distributions.  On the other hand, 
the Soffer's inequality is an inequality for {\it each quark and anti-quark
flavor}\,\cite{GJJ}.  This does not necessarily leads to the inequality
for the {\it valence} distributions.  
(The assumption in \cite{MSSV} is a sufficient condition to
guarantee the inequality for each quark and anti-quark flavor.)  
Therefore the authors of
\cite{MSSV} imposed a stronger constraint 
on $h_1$ than required by the
Soffer's inequality and their estimate of $A_{TT}$
can not be taken as the upper bound
of $A_{TT}$ solely due to the Soffer's inequality.  
Numerically, their estimate
on $A_{TT}$ is of the same order as those
in Figs. 2 and 4 (also the one in \cite{BCD}).

To summarize, we presented a first estimate of the
longitudinal-transverse spin asymmetry $A_{LT}$ 
for the polarized Drell-Yan process at RHIC and HERA-$\vec{N}$ energies
in comparison with $A_{LL}$ and $A_{TT}$.  
$A_{LT}$ normalized by the asymmetry in the parton level 
turned out to be approximately five to ten times smaller than
the corresponding $A_{TT}$, although the prediction on its absolute 
magnitude suffers from the uncertainty of the distributions, 
in particular, 
of $h_1$ as was the case for $A_{TT}$.  
The purely twist-3 contribution to $g_T$ and $h_L$ was modeled
by the bag model, and it turned out its effect on 
$A_{LT}$ is negligible compared with
the Wandzura-Wilczek contribution to $g_T$ and $h_L$.

\begin{figure}[h]
\epsfile{file=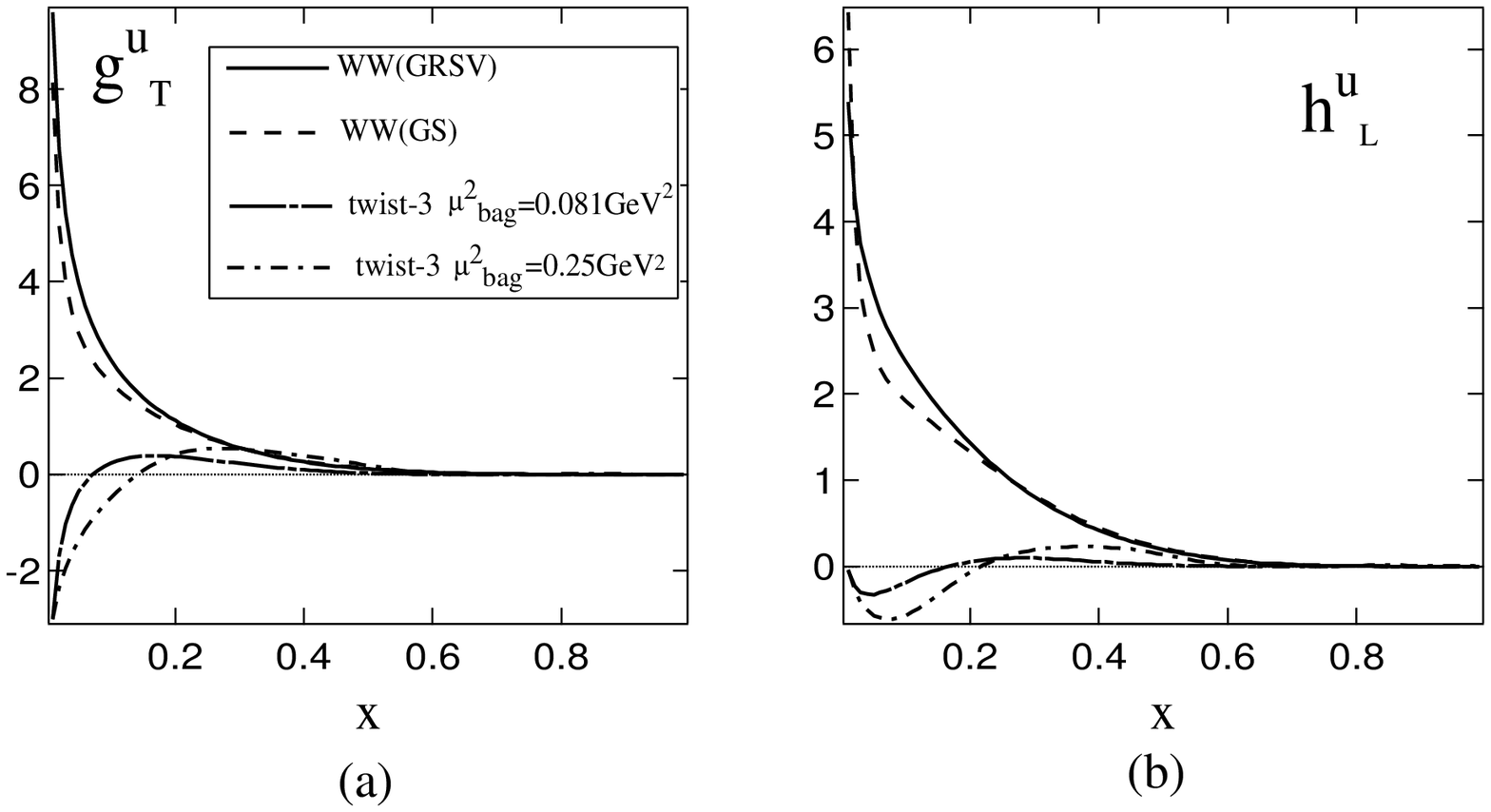,scale=0.7}
\caption[Fig. 1]{(a) $g_T^{WW}(x,\mu^2)$ obtained from the GRSV distribution
(solid line) and the GS distribution (dashed line) at $\mu^2$=1 GeV$^2$, and  
$\widetilde{g}_T(x,\mu^2)$ obtained from the bag model
calculation at $\mu^2$=1 GeV$^2$ assuming the bag scale is
$\mu_{bag}^2=0.081$ GeV$^2$ (long dash-dot line) and 
$\mu_{bag}^2=0.25$ GeV$^2$ (short dash-dot line).
(b) $h_L^{WW}(x,\mu^2)$ and $\widetilde{h}_L(x,\mu^2)$ at $\mu^2$=1 GeV$^2$.
The meaning of the lines is the same as (a).}
\end{figure}

\begin{figure}[h]
\epsfile{file=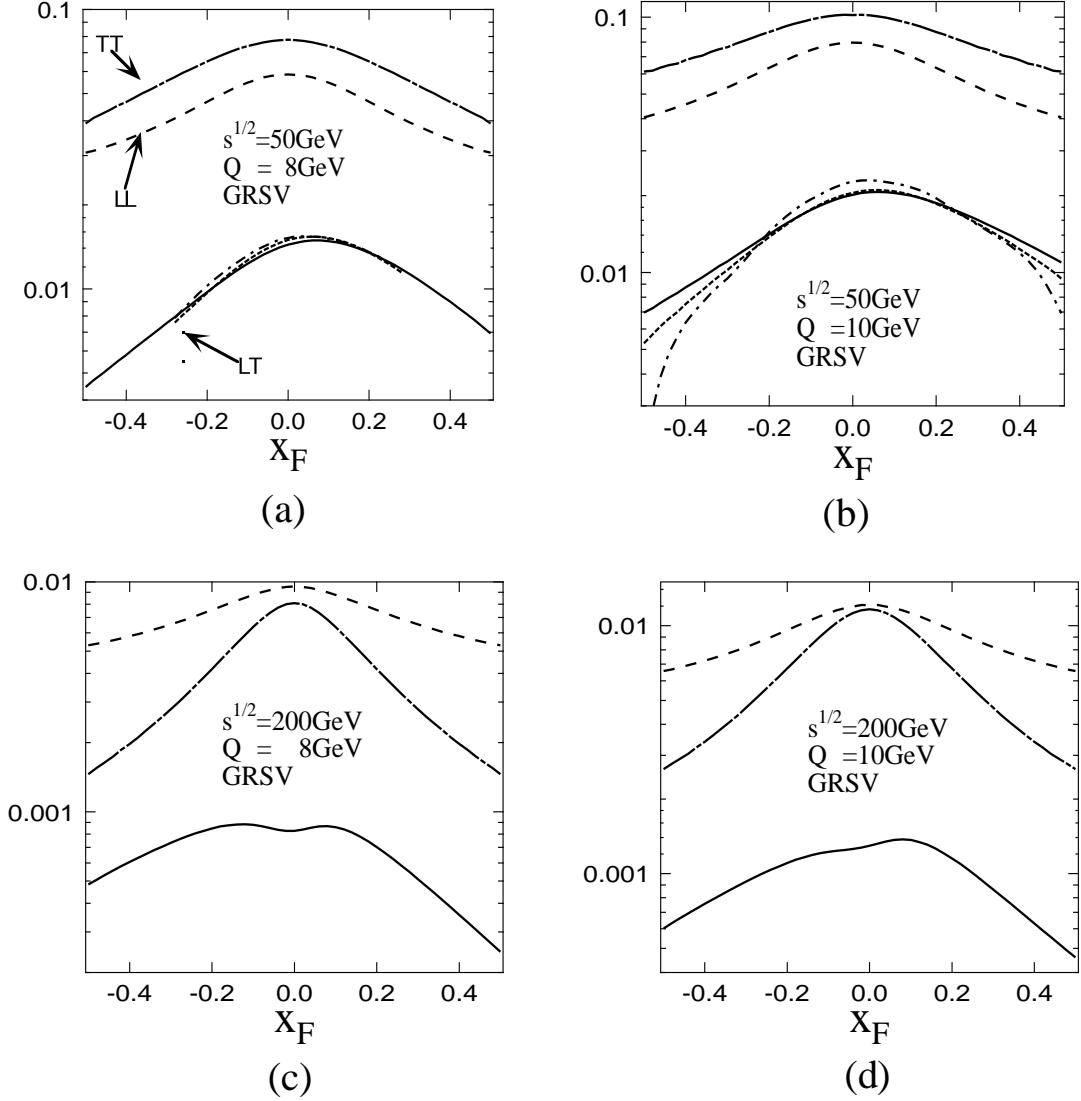,scale=0.85}
\caption[Fig. 2]{Double spin asymmetries, $\widetilde{A}_{LL}$,
$\widetilde{A}_{TT}$, $\widetilde{A}_{LT}$, 
for the polarized Drell-Yan
using the GRSV parton distribution and the bag model at
$Q=8,\ 10$ GeV and $\sqrt{s}=50,\ 200$ GeV.
The solid line denotes $\widetilde{A}_{LT}$ with only the
Wandzura-Wilczek contributions in $g_T$ and $h_L$. 
The short dash-dot line denotes $\widetilde{A}_{LT}$ with 
the bag scale
$\mu_{bag}^2=0.25$ GeV$^2$, and the dotted line
denotes
$\widetilde{A}_{LT}$ with the bag scale
$\mu_{bag}^2=0.081$ GeV$^2$.
The long dashed line corresponds to $\widetilde{A}_{LL}$, 
and the long dash-dot line corresponds to $\widetilde{A}_{TT}$.}
\end{figure}

\begin{figure}[h]
\epsfile{file=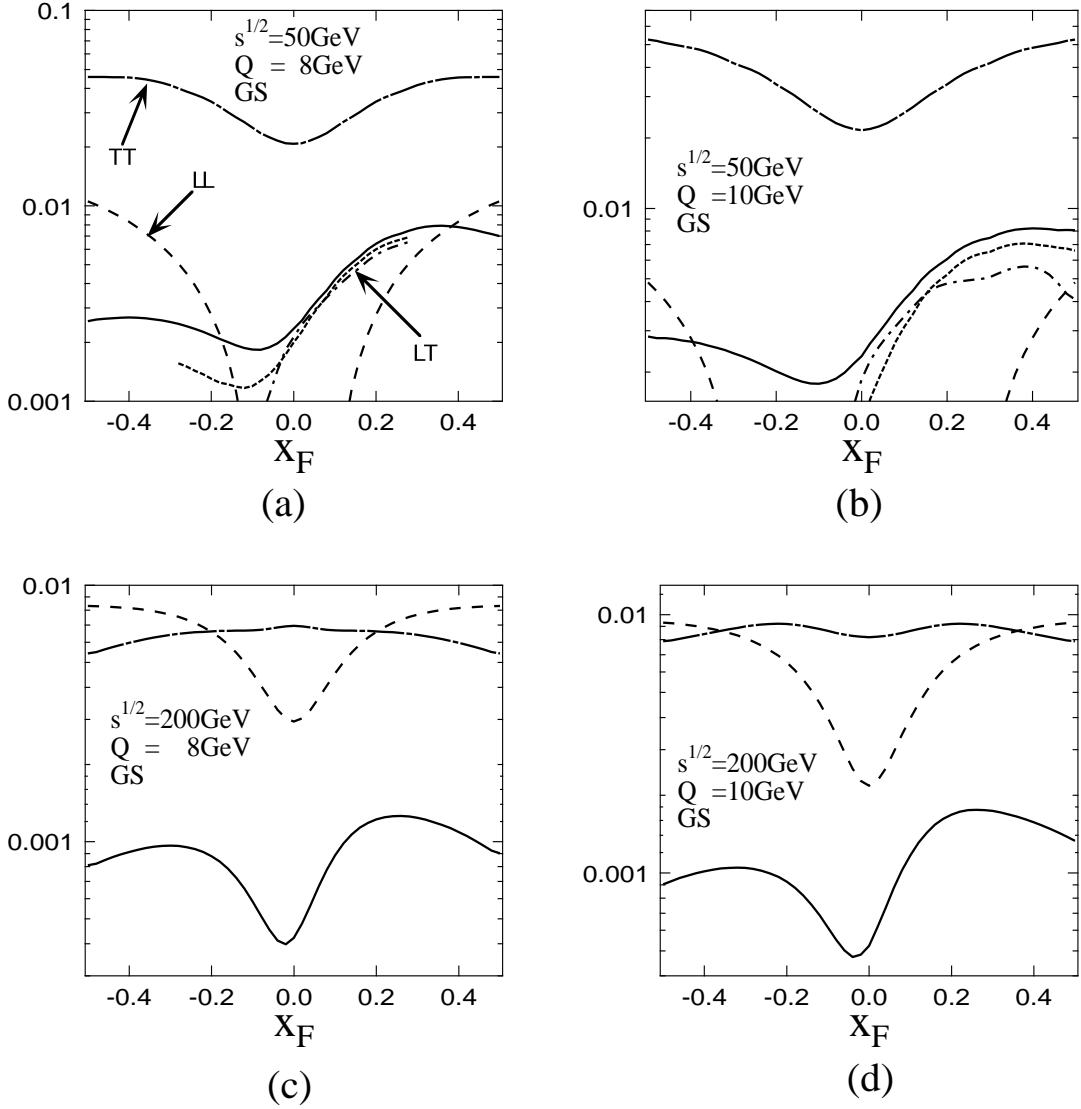,scale=0.85}
\caption[Fig. 3]{Double spin asymmetries, $\widetilde{A}_{LL}$,
$\widetilde{A}_{TT}$, $\widetilde{A}_{LT}$, for the polarized Drell-Yan
using the GS parton distribution and the bag model at
$Q=8,\ 10$ GeV and $\sqrt{s}=50,\ 200$ GeV.
The meaning of the lines is the same as Fig. 2.}
\end{figure}

\begin{figure}[h]
\epsfile{file=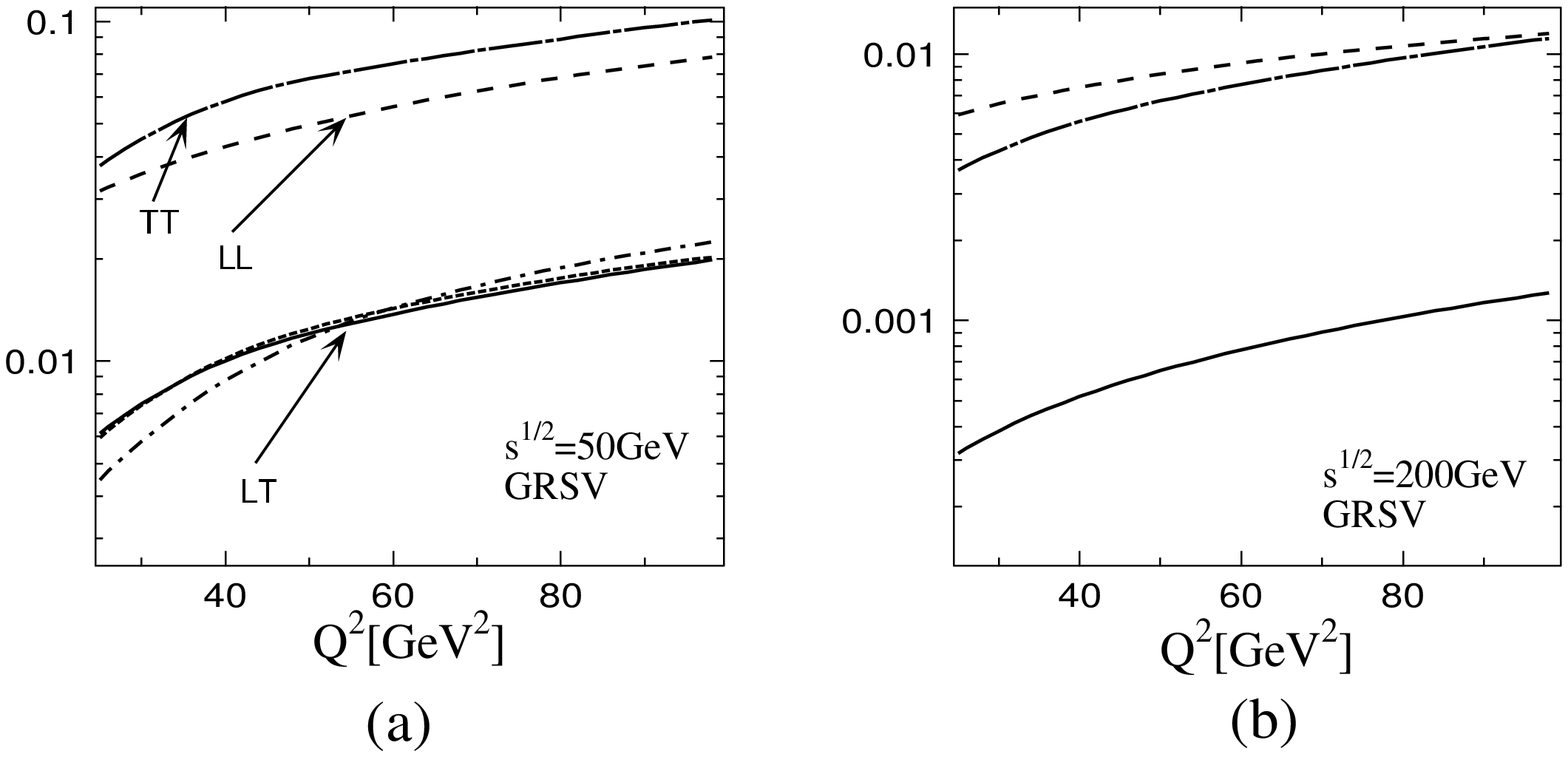,scale=0.7}
\caption[Fig. 4]{The $Q^2$ dependence of the asymmetries, $\widetilde{A}_{LL}$,
$\widetilde{A}_{TT}$, $\widetilde{A}_{LT}$, at $x_F=0$ in 
the polarized Drell-Yan at $\sqrt{s}=50$ GeV (a) and 200 GeV (b)
using the 
GRSV parton distribution and the bag model.  The meaning of the lines
is the same as Figs. 2.}
\end{figure}

\begin{figure}[h]
\epsfile{file=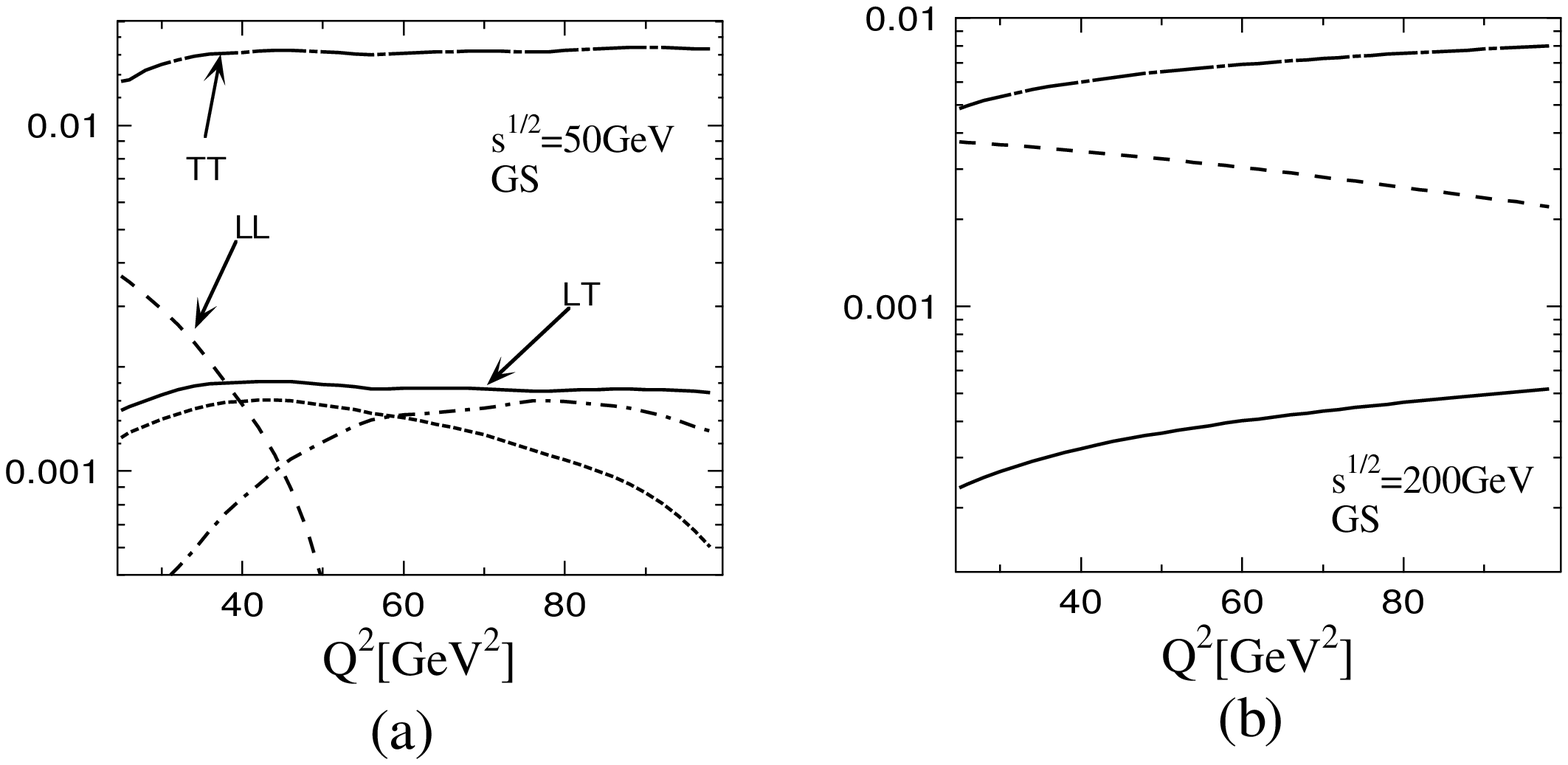,scale=0.7}
\caption[Fig. 5]{The $Q^2$ dependence of the asymmetries, $\widetilde{A}_{LL}$,
$\widetilde{A}_{TT}$, $\widetilde{A}_{LT}$, at $x_F=0$ in 
the polarized Drell-Yan at $\sqrt{s}=50$ GeV (a) and 200 GeV (b)
using 
the GS parton distribution and the bag model.  The meaning of the lines
is the same as Figs. 2.}
\end{figure}

\vspace{0.5cm}

\centerline{\bf Acknowledgement}

We thank
N. Saito for useful discussions on the RHIC kinematics
and T. Gehrmann for providing us with the Fortran code of
their parton distribution.


\begin{thebibliography}{99}

\bibitem{RS} J.P. Ralston and D.E. Soper, Nucl. Phys. {\bf B152},
109 (1979).

\bibitem{AM} X. Artru and M. Mekhfi, 
Z. Phys. {\bf C45} 669 (1990).

\bibitem{JJ} R.L. Jaffe and X. Ji, Nucl.\ Phys.\ {\bf B375}, 527 (1992).

\bibitem{CPR} J.L. Cortes, B. Pire and J.P. Ralston,
Z. Phys. {\bf C55}, 409 (1992).

\bibitem{VW} W. Vogelsang and A. Weber, Phys. Rev. {\bf D48}, 2073 (1993).

\bibitem{Kama96} B. Kamal, Phys. Rev. {\bf D53}, 1142 (1996).

\bibitem{BCD} V. Barone, T. Calarco and A. Drago, Phys. Lett. {\bf B390}, 287
(1997); Phys. Rev. {\bf D56}, 527 (1997).

\bibitem{Geh} T. Gehrmann, Nucl. Phys. {\bf B498}, 245 (1997).

\bibitem{Kama97} B. Kamal, hep-ph/9710374.

\bibitem{MSSV} O. Martin, A. Sch\"{a}fer, M. Stratmann, and W. Vogelsang,
Phys. Rev. {\bf D57}, 3084 (1998). 

\bibitem{Jg2} R.L. Jaffe, Comm. Nucl. Part. Phys. {\bf 19}, 239 (1990).

\bibitem{QS} J. Qiu and G. Sterman, Phys. Rev. Lett. {\bf 67}, 2264 (1991);
Nucl. Phys. {\bf B378}, 52 (1992). 

\bibitem{GRV} M. Gl\"{u}ck, E. Reya and A. Vogt, Z. Phys. {\bf C67},
433 (1995). 

\bibitem{MRS} A.D. Martin, R.G. Roberts and W.J. Stirling, Phys. Lett.
{\bf B354}, 155 (1995). 

\bibitem{CTEQ} H.L. Lai, et. al. Phys. Rev. {\bf D51}, 4763 (1995).

\bibitem{GRSV} M. Gl\"{u}ck, E. Reya, M. Stratmann and W. Vogelsang,
Phys. Rev. {\bf D53}, 4775 (1996).

\bibitem{GS} T. Gehrmann and W.J. Stirling, 
Phys. Rev. {\bf D53}, 6100 (1996).
 
\bibitem{ABFR} G. Altarelli, R. Ball, S. Forte and G. Ridolfi, Nucl. Phys.
{\bf B496}, 337 (1997).

\bibitem{PP} 
P.V. Pobylitsa and M.V. Polyakov, Phys. Lett. {\bf B389}, 350 (1996).

\bibitem{BBKT}
I.I. Balitsky, V.M. Braun, Y. Koike and K. Tanaka,
Phys. Rev. Lett. {\bf 77}, 3078 (1996).
See also Y. Koike and N. Nishiyama, 
Phys. Rev. {\bf D55}, 3068 (1997).

\bibitem{g2} 
A.P. Bukhvostov, E.A. Kuraev and L.N. Lipatov, 
Sov. Phys. JETP {\bf 60}, 22 (1984);\\
P.G. Ratcliffe, Nucl. Phys. {\bf B264}, 493 (1986);\\
I.I. Balitsky and V.M. Braun, Nucl. Phys. {\bf B311}, 541 (1988/89);\\
X. Ji and C. Chou, Phys. Rev {\bf D42}, 3637 (1990);\\
D. M\"{u}ller, Phys. Lett. {\bf B407}, 314 (1997);\\
J. Kodaira, Y. Yasui, K. Tanaka and T. Uematsu, Phys. Lett. {\bf B387},
855 (1996);\\
J. Kodaira, T. Nasuno, H. Tochimura, K.Tanaka, Y. Yasui,
hep-ph/9712395, Prog. Theor. Phys. in press.

\bibitem{KT} Y. Koike and K. Tanaka, Phys. Rev. {\bf D51}, 6125 (1995);\\
A.V. Belitsky and D. M\"{u}ller, Nucl. Phys. {\bf B503}, 279 (1997).

\bibitem{ABH}
A. Ali, V.M. Braun and G. Hiller,  Phys. Lett. 
{\bf B266}, 117 (1991).

\bibitem{KK} Y. Kanazawa and Y. Koike, Phys. Lett. {\bf B403}, 357 (1997). 

\bibitem{St} M. Stratmann, Z. Phys. {\bf C60}, 763 (1993);\\
X. Song, Phys. Rev. {\bf D54}, 1955 (1996).  

\bibitem{Jaffe75} R.L. Jaffe, Phys. Rev. {\bf D11}, 1953 (1975).

\bibitem{Soffer} J. Soffer, Phys. Rev. Lett. {\bf 74}, 1292 (1995).  

\bibitem{GJJ} G.R. Goldstein, R.L. Jaffe and X. Ji, Phys. Rev. 
{\bf D52}, 5006 (1995). 

\end{thebibliography}
\end{document}